\documentclass[%
 aip,
 amsmath,amssymb,
 reprint,%
 floatfix,
]{revtex4-1}

\usepackage{graphicx}
\usepackage{dcolumn}
\usepackage{bm}

\usepackage[utf8]{inputenc}
\usepackage[T1]{fontenc}
\usepackage{mathptmx}
\usepackage{etoolbox}
\usepackage{comment}

\newcommand{\rsup}{\raisebox{0.55ex}{\scalebox{0.75}{2}}}
\newcommand{\rscan}{r\rsup SCAN}
\newcommand{\bn}{bn-r\rsup SCAN}
\newcommand{\fx}{f_\text{x}}
\newcommand{\fc}{f_\text{c}}

\makeatletter
\def\@email#1#2{%
 \endgroup
 \patchcmd{\titleblock@produce}
  {\frontmatter@RRAPformat}
  {\frontmatter@RRAPformat{\produce@RRAP{*#1\href{mailto:#2}{#2}}}\frontmatter@RRAPformat}
  {}{}
}%
\makeatother
\begin{document}

\title{Balanced and broadly-normed meta-generalized gradient approximation}
\author{Yanyong Wang}
\author{Chandra Shahi}
\author{John P. Perdew}
\author{Adrienn Ruzsinszky}
 \email{aruzsin@tulane.edu}
\affiliation{Department of Physics and Engineering Physics, Tulane University, New Orleans, Louisiana 70118, USA}
\begin{abstract}
Conventional semi-local density functional approximations for the exchange-correlation energy systematically underestimate fundamental band gaps. 
Advanced meta-generalized gradient approximations (meta-GGAs) can partially overcome this limitation through their orbital dependence via the kinetic energy density. 
In principle, the flexible meta-GGA form should allow for balanced performance across diverse properties. 
However, atomization energies, reaction barrier heights, and lattice constants, as well as band gaps, exhibit different sensitivities to the large parameter space of meta-GGA functionals. 
In this work, we explore the space of appropriate norms that guide the \rscan-type functional construction through the interpolation function connecting two fundamental paradigms: the one-electron limit and the uniform electron gas limit. 
We introduce local modifications to a strongly-smoothed interpolation function and investigate their impact on functional performance. 
To further probe the limits of the meta-GGA framework, we include spin-unpolarized bonded systems among the set of norms. 
The resulting bonded-norm \bn\ meta-GGA achieves balance:
band-gap predictions approaching the accuracy of the ultra-nonlocal LAK meta-GGA and lattice constants of nearly \rscan\ accuracy.
\end{abstract}

\maketitle

\section{Introduction}

Density functional theory (DFT), which in principle exactifies the Hartree approximation for the ground-state energy and density of a many-electron system, is a powerful and efficient tool widely used in materials research and design.\cite{KohnSham1965,Jones2015} 
The accuracy and efficiency of DFT depend on the approximated exchange-correlation (xc) energy term in the Kohn--Sham equations.\cite{KohnSham1965} 
Exchange-correlation approximations, or density functional approximations (DFAs), range from the local density approximation (LDA)\cite{KohnSham1965} and generalized gradient approximation (GGA)\cite{PBE1996} to meta-generalized gradient approximations
(meta-GGAs),\cite{Sun2015,Furness2020} achieving progressively higher accuracy through the inclusion of additional ingredients. 
One of the best-known limitations of such semi-local xc approximations is the underestimation of the fundamental band gaps of solids, particularly wide-gap insulators. 
This problem is closely related to the discontinuity of the exact Kohn--Sham xc potential at integer electron numbers\cite{MoriSanchez2008Localization,
Ruzsinszky2006Fractional} and to the self-interaction error (SIE), also known as the delocalization error, in approximate xc functionals.\cite{PerdewParrLevyBalduz1982,MoriSanchez2006SIE}

The hierarchy of increasing accuracy is well represented by the so-called Jacob's ladder of density functional approximations.\cite{PerdewSchmidt2001}
Meta-GGAs occupy the third rung of Jacob's ladder,\cite{KohnSham1965,Jones2015,PBE1996,Sun2015,Furness2020} incorporating the kinetic energy density as an additional ingredient beyond those used on the first and second rungs. 
Through their dependence on the kinetic energy density, meta-GGAs are implicit functionals of the electron density, but they remain semi-local in a computational sense:
the exchange-correlation energy can be expressed as a single spatial integral because the Kohn--Sham kinetic energy density, $\tau_\sigma$,\cite{Yang2016} is available at every point in space during a self-consistent Kohn--Sham calculation. 
Good orbital-dependent functionals such as meta-GGAs and hybrid functionals are expected to give total energies whose differences often approximate\cite{YangAyers2024} one-particle-like excitation energies, while their orbital energies and band gaps in solids also approximate\cite{Perdew2017} such excitation energies.

Orbital-dependent density functionals, including meta-GGAs\cite{Becke1983Exchange,Becke1994Thermochemical} and hybrids of GGAs with a fraction of exact exchange,\cite{Becke1993Hybrid} came out of the rich imagination and insight of Axel Becke.

The Strongly Constrained and Appropriately Normed (SCAN) meta-GGA,\cite{Sun2015} proposed in 2015, was constructed to respect all 17 known exact constraints that a meta-GGA can satisfy. 
SCAN\cite{Sun2015} and the regularized and restored SCAN (\rscan),\cite{Furness2020} as orbital-dependent functionals, are conventionally implemented in a generalized Kohn--Sham (GKS) scheme,\cite{Perdew2017} in which their exchange-correlation potentials are differential operators. 
These meta-GGAs modestly increase and improve the fundamental gaps of most semiconductors, in comparison to LSDA and GGA.

These recent nonempirical meta-GGAs exploit the flexibility given by the kinetic energy density in the energy functional\cite{KohnSham1965,Jones2015} to satisfy many additional exact constraints.\cite{Perdew2005Norms,Kaplan2023Constraints,Perdew2014Gedanken} 
Constraints can be imposed on the exchange-correlation energy density or on its underlying hole density.
One constraint satisfied by these meta-GGAs is the self-interaction-free correlation energy, which properly vanishes for all one-electron densities. 
The meta-GGA exchange energy cannot be exact for all one-electron densities, but it is accurate for sufficiently compact ones. 
Recent work\cite{Shahi2026SIC} suggests that, while all exact constraints are important, the most important ones are exactness for all uniform densities and exactness for all one-electron densities. 
Of these two, only the uniform-density constraint is satisfied by most approximations.

SCAN depends upon the kinetic energy density through a dimensionless parameter $\alpha$ that can recognize and provide an appropriately different GGA-like description for each of three kinds of bonding regions: 
single covalent, metallic, and intermediate-range van der Waals.\cite{Sun2013} 
Because of the flexibility of a function of six variables, SCAN was also fitted to appropriate norms,\cite{Burke2016Locality,Schwinger1981} non-bonded systems such as atoms, in which it can be accurate for the exchange and correlation energies separately, and not just for their sum as in bonded systems. 
The restriction to non-bonded norms was important at an early stage to make the meta-GGAs genuinely predictive for bonded systems, but at the current stage bonded norms (bn's) can be added for fine-tuning. 
Long-range dispersion corrections are available and are essential for the interlayer interactions in layered materials.\cite{Wang2026Rotational}

A major obstacle to the wider use of SCAN is its numerical instability.
This problem was successfully addressed by a smooth interpolation function in rSCAN,\cite{BartokYates2019} at the cost of violating some exact constraints.
These constraints were later restored in the \rscan\ meta-GGA, which retains the accuracy of SCAN.

Recently, Lebeda, Aschebrock, and K\"ummel (LAK)\cite{Lebeda2024} exploited an additional degree of freedom in the energy density of the second-order gradient expansion for exchange and correlation to design the LAK meta-GGA, which surpasses \rscan\ in several respects. 
Despite similarities in their functional forms, LAK differs from SCAN in a number of important details. 
Nevertheless, LAK still employs a function of $\alpha$ to interpolate between the iso-orbital and slowly varying density limits. 
Like SCAN, LAK satisfies all 17 exact constraints, while being fitted to a smaller set of appropriate norms.

By virtue of its ingredients and satisfaction of exact constraints, the meta-GGA rung deserves further exploration. 
Meta-GGAs have already demonstrated remarkable achievements, including band-gap predictions whose accuracy can compete with that of the widely used HSE06 hybrid functional.\cite{HeydScuseria2004,Krukau2006}
At a computational cost only 3--6 times that of PBE, SCAN performs significantly better than the Perdew--Burke--Ernzerhof (PBE) GGA for defects in semiconductors,\cite{Sun2016Diverse} surface properties of metals,\cite{Patra2017Surfaces} the seven phases of ice,\cite{Sun2016Diverse} liquid water,\cite{Chen2017Water} liquid and supercooled silicon,\cite{Remsing2017Silicon} subtle structural distortions in ferroelectrics,\cite{Paul2017Ferroelectric,Zhang2017Ferroelectric} formation energies and structural predictions for solids, including transition-metal oxides,\cite{Yang2018Stability} and critical pressures for structural phase transitions in semiconductors.\cite{Sun2016Diverse}

Many electronic systems, including some that are intractable for simpler functionals, are now well described by more advanced meta-GGA functionals.\cite{Wang2026Rotational} 
But no universal functional is both accurate and affordable for all systems.
Reaction paths,\cite{Schmalz2024Reaction} potential energy surfaces, and conical intersections\cite{MeiYang2019} need to be described correctly.
Some transition-metal and rare-earth compounds in particular pose challenges because of strong correlation and self-interaction errors.\cite{Johansson1995Cerium,Giri2025Cerium,Gagliardi2017MCPDFT,Schafer2021Cerium,Skorodumova2001Cerium,Wen2013Actinide,Jiang2009Lanthanide} 
Charge-transfer errors are possible whenever systems interface, as when a molecule is adsorbed on a metal surface.\cite{Patra2019CO} 
Donor-acceptor systems like TTF-TCNQ are especially challenging.\cite{Caruso2014Charge} 
Heterogeneous catalysis at defect sites on exterior or interior surfaces of solids requires large supercells and correct polaronic\cite{Sah2024Polarons} or non-polaronic charge transfer.

The reduction of the delocalization error is often accompanied by a deterioration of equilibrium properties, as evidenced by the less accurate lattice constants predicted by the LAK meta-GGA relative to SCAN and \rscan. 
Nevertheless, the exact constraints satisfied by SCAN and \rscan\ are fundamental physical requirements and should not, in principle, hinder further enhancement of band gaps.

Here, we explore the meta-GGA design space by leveraging bonded norms. 
The resulting functional achieves a balanced description, combining the accurate equilibrium properties of \rscan\ with the improved band-gap predictions of the ultra-nonlocal LAK meta-GGA. 
The remainder of this paper is organized as follows. 
Section~\ref{sec:theo} presents the general meta-GGA form, the new interpolation function, and the fitting of its parameters to the band gaps and lattice constants of real solids --- the two properties we aim to balance. 
Section~\ref{sec:results} then reports the results: the structure of the interpolation functions and enhancement factor, the performance of the new functional on molecular energetics, and the band gaps and lattice constants of solids. Section~\ref{sec:summary} summarizes our conclusions.

\section{Theoretical background and functional form}
\label{sec:theo}
\subsection{Functional form}
The exchange-correlation energy of a density functional approximation is composed of an exchange part, conveniently written as the local-density exchange scaled by a dimensionless enhancement factor $F_\text{x}$, and a correlation part, with an energy per electron $\varepsilon_\text{c}$,
\begin{align}
E_\text{x}[ n_\uparrow,\, n_\downarrow] &= \frac{1}{2} \left\{ E_\text{x}[2n_\uparrow] + E_\text{x}[2n_\downarrow] \right\}, \\
E_\text{x}[n] &= \int d\mathbf{r}\, n\,\varepsilon_\text{x}^\text{LDA}(n)\,
F_\text{x}(s,\alpha), \\
E_\text{c}[ n_\uparrow,\, n_\downarrow] &=
\int d\mathbf{r}\, n\,\varepsilon_\text{c}(n_\uparrow,n_\downarrow, \vert\nabla n\vert,\alpha),
\end{align}
where $\varepsilon_\text{x}^\text{LDA}(n)=-(3/4)(3n/\pi)^{1/3}$ is the exchange energy per electron of the uniform gas, extended to spin
polarization by the exact spin-scaling relation, and $s=|\nabla n|/[2(3\pi^{2})^{1/3}n^{4/3}]$ is the reduced density gradient. 
The one ingredient beyond the (spin-)density and its gradient --- the variable that promotes the form from a GGA to a meta-GGA --- is the iso-orbital indicator $\alpha$ built from the kinetic energy density $\tau$ for the spin-unpolarized electron density $n$. 
Its definition varies across different meta-GGAs, and the form used here is the regularized definition of \rscan\cite{Sun2015,Furness2020} (adopted for the numerical robustness of the resulting potential),
\begin{equation}
\alpha=\frac{\tau-\tau_\text{W}}{\tau_\text{U}+\eta\,\tau_\text{W}},
\label{eq:alpha}
\end{equation}
where $\tau_\text{W}$ and $\tau_\text{U}$ are the von Weizs\"acker and uniform-gas kinetic energy densities, with $\eta=0.001$ chosen to keep $\alpha$ finite as the density decays to zero. 
The indicator $\alpha$ distinguishes single-orbital ($\alpha=0$), slowly varying ($\alpha\approx1$), and weakly overlapping ($\alpha\gg1$) density regions.

The limits $\alpha=0$ and $\alpha\approx1$ are described by well-chosen GGAs, and an interpolation function of $\alpha$ interpolates and extrapolates to all other values of $\alpha$:
\begin{equation}
F_\text{x}(s,\alpha) =\left[h_\text{x}^{1}(s)
+\fx(\alpha) \left(h_\text{x}^{0}-h_\text{x}^{1}(s)\right)\right]g_\text{x}(s),
\label{eq:fx-role}
\end{equation}
where $h_\text{x}^{0}$ is the exchange enhancement in the single-orbital limit, a constant pinned by the tight exchange bound for
two-electron systems; $h_\text{x}^{1}(s)$ is that in the slowly varying limit, recovering the gradient expansion for exchange; and $g_\text{x}(s)$ is a damping factor enforcing the exact
large-$s$ behavior. 
The correlation energy per electron is built analogously, its interpolation function $\fc(\alpha)$ joining the correlation energies of the same two limits. 
The interpolation function is thus the sole route by which $\alpha$ enters: it interpolates between the two descriptions in the bonding regions ($\alpha\leq1$) and extrapolates into the weakly overlapping region, which no exact condition constrains. 
A proper interpolation function must satisfy $f(0)=1$ and $f(1)=0$, so that the two GGA limits are recovered where they apply; beyond these two conditions, it is usually required to be smooth enough for a numerically stable potential, to be free of spurious oscillations between and beyond the pinned points, and to approach a finite asymptote as $\alpha\to\infty$.
\begin{table}
\caption{\label{tab:params}Fitted parameter values of \bn\ for exchange and
correlation: the coefficients of the interpolation function,
Eq.~\eqref{eq:interp}, and the width $d_{p2}$ of the gradient-expansion
damping.}
\begin{ruledtabular}
\begin{tabular}{lcccccc}
Channel & $a$ & $b$ & $c$ & $d$ & $\lambda$ & $d_{p2}$\\
\hline
Exchange    & $-$0.27          & 1.588 & 0.97741935 & \phantom{$-$}0.64
            & 0.2 & 0.94 \\
Correlation & \phantom{$-$}1.7 & 1.2   & 2.28571429 & $-$3
            & 0.5 & 0.35 \\
\end{tabular}
\end{ruledtabular}
\end{table}
\begin{table}
\caption{\label{tab:norms}Absolute appropriate-norm errors (\%) of \bn, \rscan\ and LAK:
jellium surface exchange-correlation energy averaged over $r_s=2,3,4,5$, rare-gas atom
$E_\text{xc}$, and the compressed Ar$_2$ interaction energy at three
internuclear separations.}
\begin{ruledtabular}
\begin{tabular}{lccc}
System & \bn & \rscan & LAK \\
\hline
Jellium surface       & 1.45 & 2.80 & --   \\
Ne    & 0.21 & 0.27 & 0.98 \\
Ar    & 0.26 & 0.11 & 1.00 \\
Kr    & 0.30 & 0.19 & 1.43 \\
Xe    & 0.47 & 0.26 & 1.49 \\
Ar$_2$ ($R=1.6$~\AA)  & 1.71 & 0.43 & 1.40 \\
Ar$_2$ ($R=1.8$~\AA)  & 2.95 & 0.41 & 2.78 \\
Ar$_2$ ($R=2.0$~\AA)  & 4.59 & 1.67 & 4.34 \\
\end{tabular}
\end{ruledtabular}
\end{table}

\subsection{Interpolation function}

Subject to these requirements, the interpolation function $f(\alpha)$ retains substantial flexibility, and its formula is almost the only difference between \bn\ and \rscan.
We adopt a single rational form for both channels,
\begin{equation}
f(\alpha)=\frac{1-\alpha^{3}}
{1+a\alpha+b\,\alpha^{2}+c\,\alpha^{3}
+d\,\alpha^{2}e^{-\lambda\alpha^{2}}},
\label{eq:interp}
\end{equation}
which satisfies $f(0)=1$ and $f(1)=0$ by construction. 
As a single analytic ratio with a nonvanishing denominator, $f$ is smooth to all orders. 
The cubic leading terms in the numerator and denominator
give the finite asymptote $f(\alpha\to\infty)=-1/c$.
The polynomial coefficients control $f(\alpha)$ in different ranges: $a$ sets the initial slope ($f'(0)=-a$), $c$ determines the asymptote, and $b$ adjusts the intermediate-$\alpha$ behavior.
The Gaussian-damped term $d\alpha^{2}e^{-\lambda\alpha^{2}}$ supplies exactly the local control that the other terms cannot: vanishing at both small and large $\alpha$, it reshapes the moderate-$\alpha$ window --- the factor $\alpha^{2}e^{-\lambda\alpha^{2}}$ peaking at $\alpha=1/\sqrt{\lambda}$, which exceeds one for both fitted $\lambda$ values --- without touching either pinned limit.
The fitted parameters use this freedom (see Table~\ref{tab:params}): in both exchange and correlation channels, $d$ takes the sign opposite to $a$, and because the numerator changes sign at $\alpha=1$, the damped term pushes the curve in opposite directions on the two sides of that point. 
The Gaussian-damped term adds a localized degree of freedom that allows the intermediate-$\alpha$ shape to be adjusted independently of the initial slope and asymptotic value.

\subsection{Fitting to band gaps and lattice constants of solids}

In \rscan, the free parameters were determined by fitting to appropriate norms: rare-gas atoms, jellium surfaces and the compressed Ar$_2$ dimer.\cite{Sun2015,Furness2020} 
\bn\ offers more freedom than \rscan: per channel (exchange or correlation), five coefficients of its interpolation function and the width $d_{p2}$ of the gradient-expansion damping\cite{Furness2020} --- twelve parameters in all.
The original norms cannot resolve so many parameters on their own, since many parameter sets yield similarly small norm errors.
Scoring the candidate parameter sets on two solid-state benchmark sets narrows the choice much further: a parameter set has to do well on both, and the two benchmarks do not reward the same choices. 
Instead of choosing the parameters by minimizing a composite error, we define for each norm an acceptable range of error and so filter a large set of candidate parameters down to a smaller set of acceptable parameters. 
The final fit is directed at the two properties whose balance is at issue, the SCBG15 set of $sp$-semiconductor band gaps\cite{Lebeda2024} and the LC20 set of simple-solid lattice constants, the bonded norms of this work.\cite{Sun2011,Csonka2009}
The procedure is straightforward: the parameter values are varied, and every candidate parameter set is scored on both benchmark sets; band gaps are calculated at fixed experimental geometries, whereas lattice constants are obtained by full structural relaxation.
On SCBG15 the accepted set nearly halves the gap mean absolute error (MAE) ($0.52\to0.28$~eV), widening all fifteen gaps; most gaps remain underestimated, but the systematic shortfall is reduced.
On LC20 it retains \rscan's accuracy in aggregate (both $0.026$~\AA\ MAE).

The fitting results indicate that both exchange and correlation channels carry the gap gain.
To be specific, the parameters that steer it are the initial slopes of the interpolation functions ($a_\text{x}$, $a_\text{c}$), the weight the exchange curve carries between the single-orbital limit and its asymptote ($b_\text{x}$), and the damping width $d^\text{x}_{p2}$. 
For $a_\text{x}$ and $a_\text{c}$ the direction is systematic: a more negative $a_\text{x}$ or a larger $a_\text{c}$ enlarges the gaps, usually at the cost of the LC20 lattice constants.

Table~\ref{tab:norms} lists the appropriate-norm errors of the accepted parameter set. 
The Ar$_2$ interaction energy is where the error departs most from \rscan's, being several times less accurate at all three separations.
That departure follows from the fitting: two of the gap-steering parameters, $a_\text{c}$ and $d^\text{x}_{p2}$, also strongly affect this energy, and \rscan's Ar$_2$ accuracy could not be retained while the gaps were enlarged.
The rare-gas-atom errors remain comparable to \rscan's, and the jellium-surface error is roughly half \rscan's.
Overall, the functional is balanced between band gaps and lattice constants and consistent with the appropriate norms, and we name it \bn\ after the bonded norms that guided the fit.

\section{Results and Analysis}
\begin{figure}
\includegraphics[width=0.80\columnwidth]%
{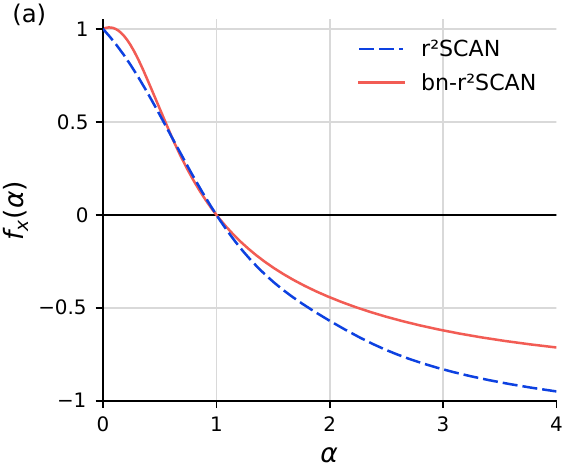}\\[6pt]
\includegraphics[width=0.80108\columnwidth]%
{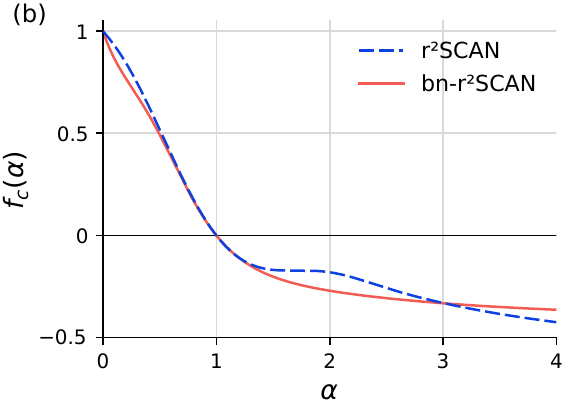}
\caption{\label{fig:interp}Interpolation functions of \bn, compared to
\rscan: (a) exchange, $\fx(\alpha)$; (b) correlation, $\fc(\alpha)$.}
\end{figure}
\label{sec:results}

We have implemented \bn\ in VASP~6.5\cite{Kresse1996,Kresse1999} and in
PySCF.\cite{Sun2018pyscf}
The solid-state test calculations use the projector augmented-wave method at a $520$~eV plane-wave cutoff, except for Cr$_2$O$_3$, which uses $700$~eV, with $\Gamma$-centered $k$-point meshes generated from \texttt{KSPACING = 0.10}~\AA$^{-1}$ and spin polarization for the transition-metal magnets in their experimental magnetic orders.
The remaining computational settings are given in the supplementary material.
Band gaps are evaluated as single points at the fixed experimental structures of 24 materials (see the supplementary material), including some non-$sp$ materials, while the separate 20-material structural set --- 14 insulators and six elemental metals --- is fully relaxed, starting from its experimental cells.
The test sets do not overlap the fitting sets. The molecular tests, as well as the rare-gas-atom and compressed Ar$_2$ calculations, were performed using PySCF.
The aug-cc-pVTZ basis set was used throughout, with a numerical integration grid of level 9.
The self-consistent field calculations were converged to an energy tolerance of $10^{-7}$~Hartree.

\subsection{The structure of the fitted interpolation functions}

Fig.~\ref{fig:interp} shows the fitted interpolation functions of \bn\ for
exchange and correlation. 
Each channel differs from its \rscan\ counterpart in a similar way: the interpolation functions remain close near $\alpha=1$ and separate on both
sides of it.
The curves meet exactly at $\alpha=1$ by construction, since both forms satisfy $f(1)=0$.
In the covalent region $\alpha\lesssim0.5$ the two
channels (exchange and correlation) move in opposite directions: 
the exchange curve leaves the single-orbital limit with a reversed, slightly positive slope and runs above \rscan's, while the correlation curve descends far more steeply and runs below. 
In the weak-overlap region ($\alpha>1$) both curves decay to less negative asymptotic values than \rscan's.
The exchange curve stays above \rscan's throughout this region, and the correlation curve rises above \rscan's beyond $\alpha\approx3$. 
The correlation curve also does not exhibit the shallow oscillation of \rscan's polynomial near $\alpha\approx2$.
Fitting to both targets together led to this shape: 
the small-$\alpha$ reshaping dominates the gap, while the large-$\alpha$ differences balance it against the lattice constants in the LC20 set.
\begin{figure}
\includegraphics[width=0.85\columnwidth]%
{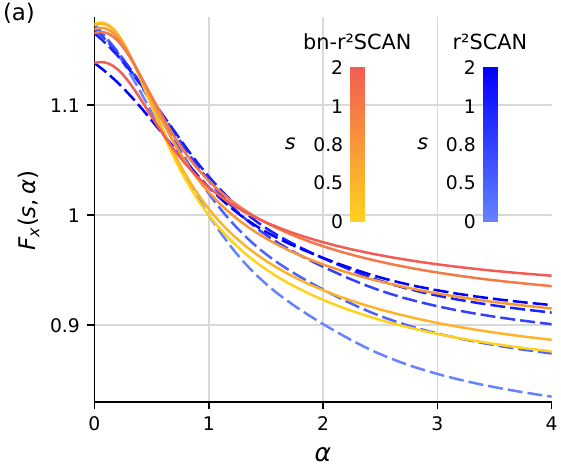}\\[6pt]
\includegraphics[width=0.85\columnwidth]%
{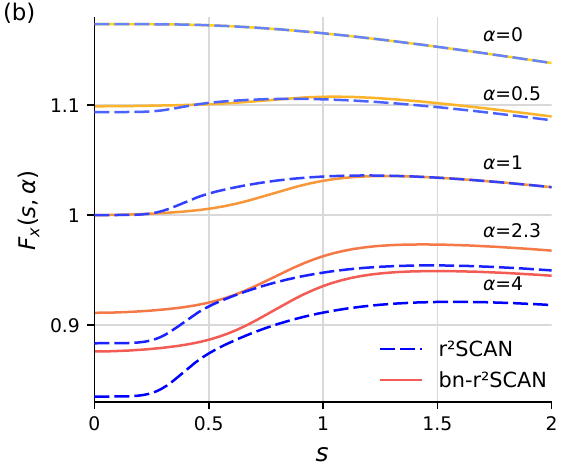}
\caption{\label{fig:fxenh}Exchange enhancement factor
$F_\text{x}(s,\alpha)$ of \bn, compared to \rscan: (a) as a function of
$\alpha$ for reduced density gradients $s=0$ (the highest curve of each color at $\alpha=0$), $0.5$, $0.8$, $1$ and $2$;
(b) as a function of $s$ for $\alpha=0$, $0.5$, $1$, $2.3$ and $4$.}
\end{figure}

Fig.~\ref{fig:fxenh} shows the exchange enhancement factor $F_\text{x}(s,\alpha)$ for a series of $\alpha$ and $s$ values. 
Its panel (a) illustrates the retained limiting behavior.
At $\alpha=0$ the \bn\ and \rscan\ curves start from the same value at each $s$ as expected, since both take the same single-orbital limit. 
At $s=0$ they cross $F_\text{x}=1$ at $\alpha=1$, the uniform-gas point. 
Beyond these points, Fig.~\ref{fig:fxenh}(a) clearly shows the pattern of $\fx$ carried into $F_\text{x}$. 
The $s$ dependence of $F_\text{x}$ in Fig.~\ref{fig:fxenh}(b) separates the two limits. 
At $\alpha=0$ the \bn\ and \rscan\ curves coincide across the whole $s$ range. 
At $\alpha=1$, where the interpolation is switched off, the difference comes from the restoration of the gradient expansion:
its coefficient follows the new $\fx$ slope, and its damping window, set by $d^\text{x}_{p2}$, is much wider, $0.94$ against \rscan's $0.361$. This gives a shallow dip of \bn\ below \rscan. 
The wider window also smooths $F_\text{x}$ along $s$, increasingly so as $\alpha$ grows, since $h^1_\text{x}$ enters Eq.~\eqref{eq:fx-role} with weight $1-\fx(\alpha)$.
Notice that beyond $\alpha\approx2.3$, \bn's
exchange enhancement factor is higher than \rscan's for all $s$, while from $\alpha\approx0.5$ to $\alpha\approx2.3$ it is lower at moderate $s$.
\subsection{Molecular energetics}
\label{sec:fintsys}
\begin{figure}[!t]
\includegraphics[width=\columnwidth]%
{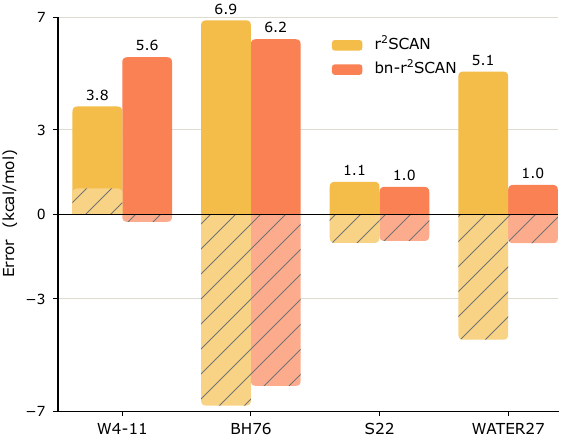}
\caption{\label{fig:gmtkn55}Mean error (ME, hatched) and mean absolute error
(MAE, solid) of \bn\ and \rscan\ on four molecular benchmark sets, in
kcal/mol.}
\end{figure}
\begin{table}
\caption{\label{tab:gmtkn55}Mean error (ME) and mean absolute error (MAE) of \bn\ and \rscan\ on the molecular benchmark sets, in kcal/mol. The two functionals differ most for WATER27.}
\begin{ruledtabular}
\begin{tabular}{lcccc}
       &\multicolumn{2}{c}{\bn}&\multicolumn{2}{c}{\rscan}\\
Subset & ME & MAE & ME & MAE \\
\hline
W4-11   & $-$0.27 & 5.58 & \phantom{$-$}0.92 & 3.83 \\
BH76    & $-$6.10 & 6.22 & $-$6.80 & 6.89 \\
S22     & $-$0.95 & 0.97 & $-$1.02 & 1.15 \\
WATER27 & $-$1.03 & 1.04 & $-$4.46 & 5.07 \\
\end{tabular}
\end{ruledtabular}
\end{table}

Molecular benchmarks probe the fit far outside its solid-state training domain. 
Four standard sets separate the main classes of molecular energetics: the W4-11 atomization energies\cite{Karton2011} for covalent bond strengths, the BH76 reaction barrier heights\cite{Zhao2005} of the GMTKN55 database,\cite{Goerigk2017} the S22 hydrogen-bonded, dispersion-bonded and
mixed dimers,\cite{Jurecka2006} and the WATER27 neutral, protonated and deprotonated water clusters.\cite{Bryantsev2009}
Table~\ref{tab:gmtkn55} lists the mean and mean absolute errors, plotted in Fig.~\ref{fig:gmtkn55} (ME hatched, MAE solid).

Both functionals underestimate the BH76 barrier heights and the S22 binding energies and overbind the WATER27 clusters. \bn, however, reduces both the mean error and the MAE in all three, and for WATER27 the MAE falls by almost a factor of five (MAE $5.07\rightarrow1.04$~kcal/mol).
A more extensive study of neutral and ionic water clusters with \bn\ is presented in Ref.~\onlinecite{Shahi2026Water}.
Detailed WATER27 results are reported in Ref.~\onlinecite{Shahi2026Water}, and not here.
Covalent atomization energies move the other way: the W4-11 MAE rises from $3.8$ to $5.6$~kcal/mol while the ME falls to near zero. 
This rise is connected to the Ar$_2$ interaction energy errors of Table~\ref{tab:norms}, as the fitting indicated. 
Overall, a parameter set chosen on solid-state properties carries over to molecular energetics, with the exact constraints still satisfied. Here too the fitting is a matter of balance.

\subsection{Band gaps and lattice constants}
\label{sec:gap_and_lc}
\begin{figure}[!t]
\noindent
\includegraphics[width=0.48\columnwidth]%
{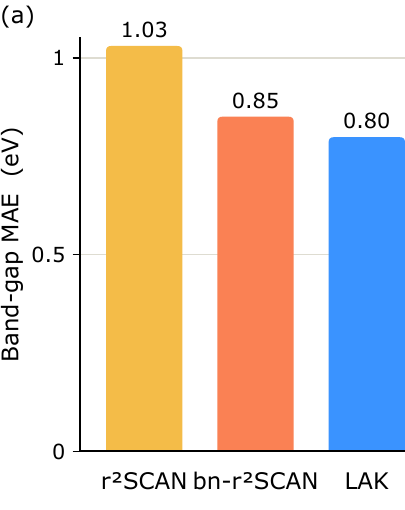}\hfill%
\includegraphics[width=0.48\columnwidth]%
{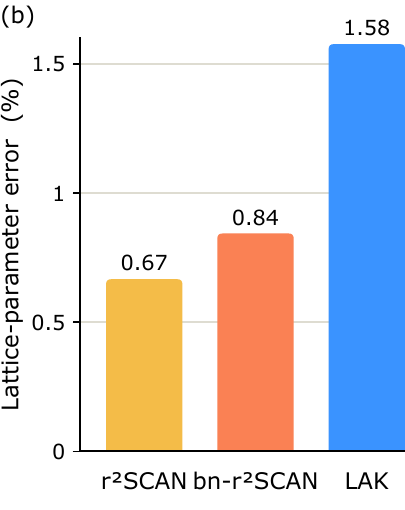}
\caption{\label{fig:solids}Accuracy of \rscan, \bn\ and LAK on solids:
(a) mean absolute error in band gaps, in eV, over 24 materials;
(b) mean absolute relative error in lattice parameters, in percent, over
20 materials.}
\end{figure}

\bn\ is assessed on solids through two complementary properties: band gaps and equilibrium lattice parameters. 
The 24-material gap set spans $0.3$--$6.4$~eV and is made up of main-group semiconductors and insulators, closed-shell $d^{0}$ and $d^{10}$ oxides, and correlated transition-metal and rare-earth oxides, the last of these being especially difficult for semi-local approximations.
Per-material gaps, relaxed cells and experimental references are given in the supplementary material.

Fig.~\ref{fig:solids}(a) compares the three functionals for the gaps, and all three underestimate them on average.
Over the 24 materials \bn\ lowers the MAE from $1.03$~eV to $0.85$~eV, approaching LAK's $0.80$~eV (see Table~\ref{tab:gaps}). 
For the 17 main-group and closed-shell-oxide materials it lies between \rscan\ and LAK, which is what the fit supplies directly, SCBG15 being a set of $sp$ semiconductors. 
The larger gains fall among the seven correlated oxides, although no oxide of this kind entered the fit. 
Cr$_2$O$_3$, Co$_3$O$_4$, CoO and NiO have absolute gap errors approximately $0.5$~eV smaller than those obtained with \rscan, and each comes closer to experiment than LAK.
The gaps decrease slightly for MnO and ZnO, by less than $0.1$~eV. InAs also becomes less accurate, its enlarged gap overshooting the experimental reference.
\begin{table}[!t]
\caption{\label{tab:gaps}Mean absolute error (MAE) and mean error (ME) of
generalized Kohn--Sham band gaps relative to experiment over the 24-material set, in eV.}
\begin{ruledtabular}
\begin{tabular}{lD{.}{.}{2.2}D{.}{.}{2.2}D{.}{.}{2.2}}
    & \multicolumn{1}{c}{\bn} & \multicolumn{1}{c}{\rscan} & \multicolumn{1}{c}{LAK} \\
\hline
MAE & 0.85 &   1.03 &  0.80 \\
ME  & -0.82 & -1.03 & -0.77 \\
\end{tabular}
\end{ruledtabular}
\end{table}

The gap accuracy of LAK comes with a structural cost that \bn\ largely avoids, as Fig.~\ref{fig:solids}(b) shows. 
The errors in the lattice parameters $a$ and $c$ are measured against experiment and averaged over the crystallographic axes, directly for the mean relative error (MRE) and in absolute value for the mean absolute relative error (MARE), first within each crystal and then over the set, so that every material carries equal weight.
This set is scored in relative rather than absolute terms because, unlike the cubic solids of LC20, it contains materials whose $a$ and $c$ differ in magnitude.
On the full structural set \bn\ stays close to \rscan, both below $1\%$, while LAK exceeds $1.5\%$. 
The class-resolved errors in Table~\ref{tab:lc} locate the difference:
\rscan\ and \bn\ separate in the 14 insulators, while for the six elemental metals the two are almost the same and the LAK MARE is approximately 1.8 times as large.

\begin{table}[!t]
\caption{\label{tab:lc}Mean absolute relative error (MARE) and mean relative
error (MRE, in parentheses) of \bn, \rscan\ and LAK on lattice parameters, in
percent, resolved by material class.}
\begin{ruledtabular}
\begin{tabular}{lccc}
Set & \bn & \rscan & LAK \\
\hline
All (20)        & 0.84 (0.63) & 0.67 (0.59) & 1.58 (1.48) \\
Insulators (14) & 0.53 (0.23) & 0.30 (0.19) & 1.03 (0.90) \\
Metals (6)      & 1.57 (1.57) & 1.53 (1.53) & 2.84 (2.84) \\
\end{tabular}
\end{ruledtabular}
\end{table}

Across the two solid-state properties, \bn\ stays near whichever of the other two functionals is the better one.
Against \rscan\ it improves band-gap accuracy substantially, the gain concentrated in the correlated oxides, with the lattice-parameter MARE rising from $0.67\%$ to $0.84\%$;
against LAK it reaches nearly the same gap accuracy at roughly half the lattice error. 
This is consistent with the reshaping of the interpolation functions and the exchange enhancement factor in Figs.~\ref{fig:interp} and \ref{fig:fxenh}, the gaps responding strongly while the relaxed structures stay near the \rscan\ baseline.

\section{Conclusion}
\label{sec:summary}
Inheriting the exact constraints of \rscan\ through its construction, \bn\ adopts a rational interpolation function with more freedom than \rscan's piecewise one, and makes use of that freedom on solids: twelve parameters fitted to band gaps and lattice constants. 
On the fitting sets, it enlarges the gaps while the lattice constants stay at \rscan's accuracy, and a similar tradeoff appears across the tested solids and molecules, although the effects depend on the property.
Over 24 semiconductors, wide-gap insulators, and oxides, the gap MAE falls from $1.03$~eV to $0.85$~eV, close to LAK's, with the largest improvements occurring for the correlated transition-metal oxides, a class absent from its fit set, while the lattice MARE of the separate 20-material structural set increases from $0.67\%$ to $0.84\%$, roughly half of LAK's.
Among molecules, reaction barrier heights, dispersion-bound dimers, and water clusters all gain, the water-cluster MAE dropping from $5.07$~kcal/mol to $1.04$~kcal/mol, whereas covalent atomization energies lose accuracy.

\rscan\ attains the better lattice constants and LAK the better gaps, each at the expense of the other property, whereas \bn\ stays near the better of the two on either property, so one self-consistent description covers both. 
The parameters that improve the gaps worsen the lattice constants, so the parameter set was accepted on the basis of its combined performance for the two properties.
The appropriate norms served as acceptance bounds rather than as terms in an objective function, and their errors stay near \rscan's for the jellium surface and the rare-gas atoms; 
this balance comes with larger errors than \rscan's for the compressed Ar$_2$ interaction energy and the covalent atomization energies.
This balance motivates future application of \bn\ to properties that require accurate electronic structure and equilibrium geometry simultaneously, such as deformation potentials, phonon-limited transport, and defect charge-transition levels.
With the exact constraints satisfied, we fitted a meta-GGA to solid-state properties and assessed it against appropriate norms, without a general loss of transferability.

\begin{acknowledgments}
The work of YW, JPP, and AR was supported by the National Science Foundation under grant no. CHE-CTMC-2533416. The work of CS was supported by the Department of Energy, Office of Science, Basic Energy Sciences, under grant no. DE-SC0018331. The work of JPP was also supported by the National Science Foundation under grant no. DMR-2426275.
YW thanks Timo Lebeda for providing inputs for the SCBG15 dataset.
\end{acknowledgments}

\section*{Author Declarations}
\subsection*{Conflict of Interest}
The authors have no conflicts to disclose.

\subsection*{Author Contributions}
\textbf{Yanyong Wang}: Data curation (equal); Formal analysis (equal);
Investigation (lead); Methodology (equal); Software (equal);
Visualization (equal);
Writing -- original draft (equal); Writing -- review \& editing (equal).
\textbf{Chandra Shahi}: Data curation (equal);
Investigation (supporting); Methodology (supporting); Software (equal);
Visualization (equal);
Writing -- review \& editing (equal).
\textbf{John P. Perdew}: Conceptualization (equal); Formal analysis (equal);
Methodology (equal);
Supervision (equal); Writing -- review \& editing (equal).
\textbf{Adrienn Ruzsinszky}: Conceptualization (equal); Formal analysis (equal);
Methodology (equal); Supervision (equal); Writing -- original draft (equal);
Writing -- review \& editing (equal).

\section*{Data Availability Statement}
The data that support the findings of this study are available within the
article and its supplementary material.

\bibliography{bn-r2scan}
\end{document}